# High pressure Ca-VI phase between 158-180 GPa: Stability, electronic structure and superconductivity

**M. Aftabuzzaman, A. K. M. A. Islam**[*]

Department of Physics, Rajshahi University, Rajshahi-6205, Bangladesh

**Abstract**

We have performed *ab initio* calculations for new high-pressure phase of Ca-VI between 158-180 GPa. The study includes elastic parameters of mono- and poly-crystalline aggregates, electronic band structure, lattice dynamics and superconductivity. The calculations show that the orthorhombic *Pnma* structure is mechanically and dynamically stable in the pressure range studied. The structure is superconducting in the entire pressure range and the calculated $T_c$ (~25K) is maximum at ~172 GPa, where the transfer of charges from $4s$ to $3d$ may be thought to be completed.

## 1. Introduction

The physical properties of some of the alkaline-earth metals like Ca, Sr, and Ba have been studied both theoretically and experimentally in view of *s-d* electron transfer under high pressure [1]. Calcium experiences a series of phase transitions under pressure. It is a face-centered cubic (fcc) (Ca-I) lattice at ambient conditions. Under compression the structural transitions are [2]: body-centered cubic (Ca-II)→simple cubic (Ca-III)→$P4_12_12$ (Ca-IV)→*Cmca* (Ca-V). Like Ba [3-6] and Sr [4-7], Ca was found to show superconductivity under pressure. Dunn and Bundy [8] first observed the superconductivity of Ca at 44 GPa as a small drop in the electrical resistance at 2 K within the sc structure. Subsequently it was demonstrated by Okada *et al.* [9] that the superconducting transition temperature $T_c \leq 3$ K above 85 GPa, which increases linearly with pressure up to 15 K at 150 GPa. Furthermore $T_c$ is experimentally reported by Yabuuchi *et al.* [10] to rise with increase in pressure in Ca-IV ($P4_12_12$) and Ca-V (*Cmca*) phases, which reaches to 25 K at 161 GPa.

A number of theoretical works on the high-pressure polymorphs of Ca have been carried out recently [11-14]. Yao *et al.* [13] have previously made theoretical calculation and reported a *Pnma* structure between 110-130 GPa which they identified as Ca-IV. Recently Ishikawa *et al*. [14] reinvestigated the phase diagram of calcium at high pressure, taking all theoretically predicted and experimentally reported structures into account. They proposed two phases Ca-VI (*Pnma,* 117-135 GPa) and Ca-VII (*I*4/*mcm*(00γ)) above 135 GPa. Additionally they found the *Pnma* phase after Ca-V (*Cmca*), which they introduced from the data of Nakamoto *et al*. [2]. This structure is same as proposed by Yao *et al*. [13] Ishikawa *et al*. [14] also remarked that Ca-V (109-117 GPa) may be realized in the narrow pressure region of less than 10 GPa and the highest superconducting $T_c$ of 25 K experimentally reported may be realized in the Ca-VI or Ca-VII phase.

---

[*] Corresponding author: azi46@ru.ac.bd

Lei *et al.* [11] carried out a theoretical study on the superconductivity of Ca within the rigid muffin-tin approximation (RMTA) and utilized the approximate formula $<\omega^2> = 0.5\theta_D^2$ without considering the distribution of the phonon density of states. This approximation yields a much larger $T_c$ compared to experiment. Gao *et al.* [15] have presented a study on the electronic behaviour, lattice dynamics and superconductivity of sc Ca up to 100 GPa. Several interesting phenomena are observed. Firstly, it is suggested that the electronic topology transitions (ETT) at the X point is responsible for the observed anomaly in electrical resistance at 40 GPa. Secondly, they predict that the sc Ca is dynamically unstable in the pressure range over its existence. This observation either suggests a need for the re-examination of the sc structure, or indicates a large anharmonic effect which stands to stabilize the phase. Lastly the observed increase of $T_c$ under pressure is mainly due to the enhanced electron-phonon coupling (EPC) matrix element and the larger electronic DOS. The latter results from the transfer of charges from 4$s$ to 3$d$ under pressure.

Very recently Nakamoto *et al.* [2] using synchrotron radiation carried out x-ray diffraction measurements of calcium at pressure up to 172 GPa at room temperature and noted a high-pressure phase "Ca-VI" above 158 GPa. An orthorhombic *Pnma* structure was determined by a Rietveld analysis. The obtained structure matched to the ones predicted by Yao *et al.* [13] well. However, their predicted pressure range did not match the experimentally observed pressure range. Nakamoto *et al.* [10] in a previous paper reported that the highest $T_c$ of 25 K in the element was observed at 161 GPa but the phase was previously considered to be Ca-V (*Cmca*). But the latest x-ray measurement indicates that the phase could possibly be a mixture of Ca-V and Ca-VI, with clear Ca-VI at 172 GPa. They suggested the necessity of phase study at low temperature and high pressure. Yin *et al.* [16], on the other hand, theoretically showed that the *Pnma* structure is favored at pressure over 140 GPa. Previously Ahuja *et al.* [17] calculated the top and bottom of the Ca 3$p$, 3$d$ and 4$s$ bands as a function of pressure and indicated that the *s*–*d* electronic transfer completes at $V/V_0 \sim 0.2$, where the 4$s$ band becomes empty. However, the structural sequence used in the calculation fcc →bcc →sc → hcp is different from that found in the experiments [2, 10]. Therefore the *s*–*d* transfer cannot be considered complete even in the Ca-IV phase and hence it seems that $T_c$ may increase above 25K on further compression above 161 GPa. Higher-pressure experiments are now in progress by Nakamoto *et al.* [2].

The structural phase transitions reported so far and the interesting superconductivity in Ca-VI (*Pnma*) have brought new avenues in front of us. The present work will therefore focus on the newly discovered phase "Ca-VI" (*Pnma*) at pressure above 158 GPa to shed further light on the results reported earlier, and particularly new results between 172-180 GPa. An analysis of the crystal structure, electronic band structures, lattice dynamics and EPC of the Ca-VI wi be made. The dynamic stability would also be an issue at hand. The calculation of elastic parameters of mono- and poly-crystalline Ca-VI in the pressure region above 158 GPa would enable us to comment on some elastic behavior as well as the mechanical stability of Ca-VI.

## 2. Method of computations

The *ab-initio* calculations were performed using DFT formalism as implemented in the CASTEP [18]. The geometrical optimization was done for high-pressure orthorhombic phase of Ca (space group 62 *Pnma*) treating the system as metallic with density mixing treatment of electrons. The generalized gradient approximation (GGA) of Perdew, Burke, and Ernzerhof for solids [19] potential has been incorporated for the simulation. We have used a 6×8×9 Monkhorst grid to sample the Brillouin zone. All structures have been fully optimized until internal stress and forces on each atom are negligible. For all relevant calculations the plane wave basis set cut-off used is 310 eV and the convergence criterion is $0.5 \times 10^{-5}$ eV/atom.

Calculations of phonon spectra, electron-phonon (e-ph) coupling and phonon density of states were performed using plane waves and pseudopotentials with QUANTUM ESPRESSO [20]. We employed ultrasoft Vanderbilt pseudopotentials [21], with a cut-off of 50 Ryd for the wave functions, and 400 Ryd for the charge densities. The *k*-space integration for the electrons was approximated by a summation over a 8×8×8 uniform grid in reciprocal space, with a Gaussian smearing of 0.02 Ryd for self-consistent cycles. Dynamical matrices and e-ph linewidths were calculated on a uniform 2×2×2 grid in phonon $q$-space. Phonon dispersions and DOS were then obtained by Fourier interpolation of the dynamical matrices, and the Eliashberg function by summing over individual linewidths and phonons.

### 3. Results and discussions

*3.1 Geometrical optimization*

The optimized structure of the calcium orthorhombic phase VI (*Pnma*) is shown in figure 1. The lattice parameters and the converged atomic positions after the optimization are given in Table I. The nearest-neighbor distances for Ca-VI at pressures 158-180 GPa have been investigated. At 172 GPa, there are fourfold $d_1$=2.2278 Å, twofold $d_2$=2.2517 Å, and twofold $d_3$=2.2529 Å, compared to corresponding values 2.268, 2.273 and 2.275 Å, respectively found experimentally by Nakamoto [2]. The next-nearest distance $d_4$=2.7554 Å is located much further than $d_3$. The coordination number for Ca-VI is thus found to be 8.

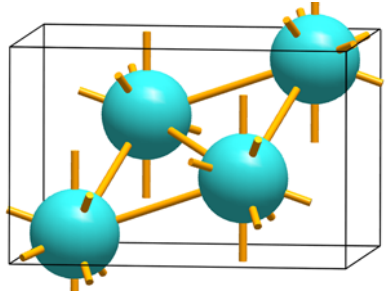

**Figure** 1. (Colour online) Unit cell of Ca-VI (*Pnma*) at 172 GPa.

**Table 1**. The optimized lattice parameters and fractional coordinates of Ca- VI (*Pnma*).

| Pressure (GPa) | $a$ (Å) | $b$ (Å) | $c$ (Å) | $V$ (Å$^3$) | Fractional coordinates | | |
|---|---|---|---|---|---|---|---|
| | | | | | $x$ | $y$ | $z$ |
| 158 | 4.2824 | 3.2390 | 2.7982 | 38.81 | 0.169654 | 0.250000 | 0.116566 |
| 161 | 4.2764 | 3.2339 | 2.7892 | 38.57 | 0.169600 | 0.250000 | 0.116680 |
| 165 | 4.2690 | 3.2280 | 2.7768 | 38.27 | 0.169491 | 0.250000 | 0.116721 |
| 172 | 4.2555 | 3.2198 | 2.7554 | 37.75 | 0.169183 | 0.250000 | 0.116283 |
| 172[a] | 4.2760 | 3.2290 | 2.8810 | 39.78 | 0.170400 | 0.250000 | 0.114800 |
| 180 | 4.2416 | 3.2124 | 2.7305 | 37.20 | 0.168789 | 0.250000 | 0.116106 |

[a]Expt [2]

*3.2 Elastic parameters of mono- and poly-crystalline Ca-VI and mechanical stability*

We performed systematic first-principles calculations of the elastic parameters of Ca-VI single crystal, such as the elastic constants $C_{ij}$, the bulk moduli $B$ and the shear moduli $G$ as a function of pressure (Table 2). These parameters are widely used for describing the elastic behavior of materials. The calculated nine independent elastic constants ($C_{11}$, $C_{12}$, $C_{13}$, $C_{23}$, $C_{22}$, $C_{33}$, $C_{44}$, $C_{55}$ and $C_{66}$) of Ca-VI at different pressures are found to satisfy the stability criteria for orthorhombic crystal [22]: $C_{ii} > 0$ ($i=1,2...6$), $C_{11} + C_{22} - 2C_{12} > 0$, $C_{11} + C_{33} - 2C_{13} > 0$, $C_{22} + C_{33} - 2 C_{23} > 0$ and $(C_{11} + C_{22} + C_{33} + 2(C_{12} + C_{13} + C_{23}) > 0$.

**Table 2**. Elastic parameters (all in GPa, except for $\nu$) for mono- and poly-crystalline Ca-VI phase.

| P (GPa) | Monocrystalline | | | | | | | | | Polycrystalline | | | |
|---|---|---|---|---|---|---|---|---|---|---|---|---|---|
| | $C_{11}$ | $C_{12}$ | $C_{13}$ | $C_{23}$ | $C_{22}$ | $C_{33}$ | $C_{44}$ | $C_{55}$ | $C_{66}$ | $B$ | $G$ | $Y$ | $\nu$ |
| 158 | 928 | 315 | 356 | 469 | 626 | 508 | 267 | 165 | 187 | 475 | 153 | 415 | 0.355 |
| 161 | 953 | 325 | 366 | 485 | 652 | 530 | 282 | 176 | 195 | 492 | 162 | 438 | 0.352 |
| 165 | 976 | 338 | 375 | 497 | 664 | 546 | 287 | 182 | 200 | 505 | 166 | 449 | 0.352 |
| 172 | 1013 | 360 | 388 | 534 | 713 | 577 | 307 | 185 | 209 | 534 | 173 | 468 | 0.354 |
| 180 | 1064 | 395 | 406 | 583 | 782 | 580 | 325 | 200 | 230 | 561 | 174 | 473 | 0.360 |

In order to study elastic anisotropy of Ca-VI we now calculate the Zener's anisotropy index $A$ [23]. For isotropic case $A = 1$, while the deviation from unity measures the degree of elastic anisotropy. The calculated values of $A$ are 0.87, 0.90, 94, and 0.97 at pressures of 158, 161, 172, and 180 GPa, respectively. It is thus seen that Ca-VI approaches towards isotropy as pressure increases.

The crystals are usually synthesized in the form of polycrystalline substances. Thus the numerical estimates of the mechanical characteristics in the polycrystalline aggregates are desirable. The Voigt-Reuss-Hill procedure [24-26] is frequently used for estimating the elastic characteristics of polycrystalline materials using $C_{ij}$-values for single crystals. Hill[25] proved that the Viogt and Reuss equations represent upper and lower limits of true polycrystalline constants. He showed that the polycrystalline moduli ($B_H$, $G_H$) are the arithmetic mean values of the moduli (monocrystalline values) in the Voigt ($B_V$, $G_V$) and Reuss ($B_R$, $G_R$) approximation. The Young's modulus $Y$ and Poisson's ratio $\nu$ are then computed from these values using the following relationship: $Y = 9BG/(3B + G)$, $\nu = (3B - Y)/6B$. The bulk moduli ($B_H \equiv B$), shear moduli ($G_H \equiv G$), Young's moduli ($Y$), and Poisson's ratios ($\nu$) in the Voigt-Reuss-Hill approximation for polycrystalline Ca-VI are presented in Table II. The expression for Reuss and Voigt moduli can be found elsewhere [27]. The calculated moduli reveal Ca-VI to be extremely hard material at the pressures under consideration.

According to Pugh [28], a given material is classified as brittle if $B/G < 1.75$. The results of our calculations give $B/G \sim 3.03 - 3.10$, which should behave as a ductile material. The Poisson's ratios of covalent systems are known to be small ($\nu \sim 0.1$), while those for ductile metallic materials $\nu$ are typically $\sim 0.33$ [29]. We can therefore assume that Ca-VI examined belong to metallic like system at all pressures.

The Debye temperatures using the formula and method given elsewhere [27] are found to be 742, 762, 771, 785, and 786 K at 158, 161, 165, 172, and 180 GPa, respectively. Unfortunately, there are no data, whether theoretical or experimental, available for comparison at the moment.

*3.3 Electronic band structure and DOS*

The calculated band structure and projected densities of states (DOS) to *s*, *p*, and *d* orbitals of the *Pnma* structure at 172 GPa are shown in figures 2 (a,b). Similar features of band structure and DOS are observed at other pressures considered here (figures not shown). In figure 2 (b) we plot the total and projected DOS for Ca in Ca-VI at 172 GPa. $E_F$ falls slightly away from the pseudogap and into the all important Ca *d* bands, which gives the value of total

DOS 1.74 states/eV. The Ca *d* contribution changes from ~ 95% to ~ 96% of total DOS for a pressure change from 158 to 180 GPa.

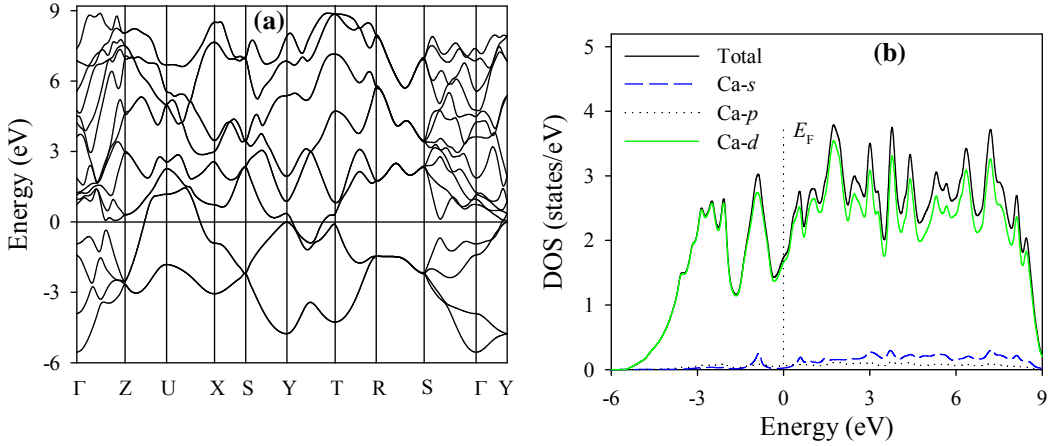

**Figure** 2. (a) Band structure, (b) (color online) total and projected DOS to *s*, *p*, and *d* orbitals of Ca-VI at 172 GPa.

*3.4 Phonon spectrum, electron-phonon coupling and superconductivity*

To investigate the occurrence of superconductivity, the phonon linewidths, EPC parameter have been calculated using perturbative linear response method within the density-functional approximation [30]. The phonon dispersion curves, and phonon density of states (PHDOS) as a function of frequency of the *Pnma* structure were evaluated at 158, 161, 165, 172 and 180 GPa (Figure shown only for 161, 172 GPa). The *Pnma* structure is found to be stable in the entire pressure range of 158-180 GPa as indicated by the lack of imaginary frequencies. The two PHDOS show that higher frequency modes exist for 172 GPa than 161 GPa. The PHDOS strength is distributed over frequency, having peaks for mid- and also high-range frequency phonons.

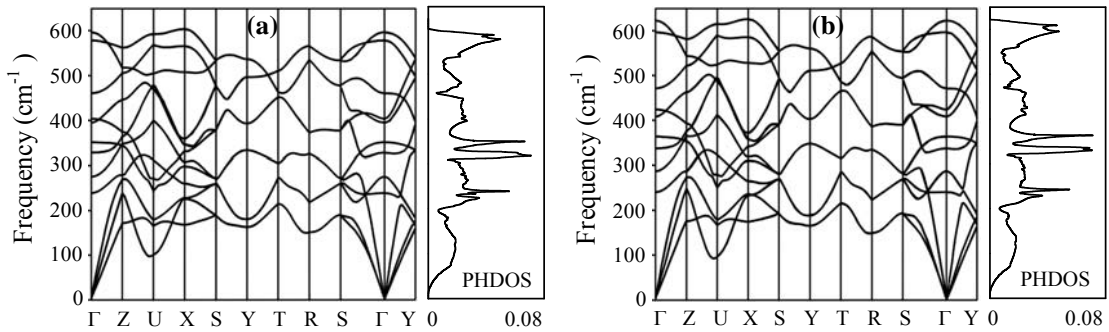

**Figure** 3. Phonon dispersion curves (*left*) and phonon DOS (*right*) at (a) 161 and (b) 172 GPa.

At 158 GPa, the calculated coupling constant $\lambda$ and phonon frequency logarithmic average $\omega_{\log}$ for the *Pnma* phase are 0.896 and 383 K, respectively. The corresponding values of the

parameters at 161, 165, 172, and 180 GPa are: 0.903, 408 K; 0.915, 405 K; 0.934, 396 K and 0887, 411 K, respectively. The superconducting critical temperature $T_c$ can be estimated from the Allen-Dynes modified McMillan equation [31]:

$$k_B T_c = \frac{\hbar \omega_{\log}}{1.2} \exp\left[-\frac{1.04(1+\lambda)}{\lambda - \mu^*(1+0.62\lambda)}\right] \quad (1)$$

Using an empirical value 0.1 for the Coulomb pseudopotential $\mu^*$, the estimated $T_c$ at 172 GPa is 24.7 K for the Ca-VI *Pnma* phase. We show a plot of calculated $T_c$ versus $P$ in figure 4. The theoretically predicted maximum value occurs at 172 GPa and is close to the experimentally observed $T_c$ of 25 K at 161 GPa [10].

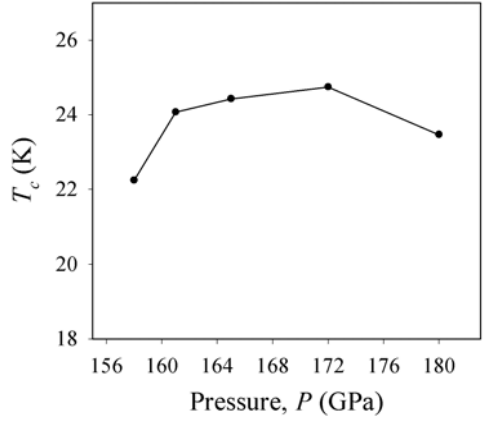

**Figure** 4. $T_c$ versus pressure of Ca-VI (*Pnma*).

In fact there is some uncertainty in the estimation of $T_c$ due to the lack of knowledge of the value of $\mu^*$ and its pressure dependence. Since the pressure dependence of $\mu^*$ is shown to be negligible [32] the only uncertainty is in the value of $\mu^*$ itself. Usually it is taken as 0.1- 0.13. Our calculated values may be slightly overestimated because of the Coulomb pseudopotential (0.1) but the trend of the resulting $T_c$ passes near the experimental value at 161 GPa. The observed high $T_c$ may be attributed to a combination of rather large $\lambda$ and that almost all phonon modes contribute to the EPC processes leading to a large $\omega_{\log}$.

The high-pressure electronic properties of Ca (see section C) can be discussed in the light of pressure-induced 4$s$-3$d$ electronic transfer [1, 17, 33]. At high pressure, the $d$-bands near the Fermi surface shift downward to points where valence electrons in Ca transfer from $s$-bands to $d$-bands. The $d$-band occupation number ($n_d$ of electrons) determines the crystal stability at large compression. This is considered to be an indication of the strength of electron-phonon interaction leading to superconductivity. In the case of Ca-VI, $T_c$ increases with pressure (with a maximum at ~172 GPa) and then it decreases. Thus the $s$–$d$ electronic transfer at 172 GPa may be thought to be completed [34].

## 4. Conclusion

The phase Ca-VI is in the *Pnma* phase which is seen to be mechanically and dynamically stable in the entire pressure range of 158-180 GPa. The calculated Zener's anisotropy index shows that Ca-VI approaches towards isotropy as pressure increases. Further from an analysis of the elastic parameters of polycrystalline aggregates of Ca-VI, the material is found to behave as a hard metallic like system for the pressure range considered here.

The superconducting critical temperatures $T_c$ have been calculated at five pressures: 158, 161, 165, 172 and 180 GPa. The theoretically predicted maximum value occurs at 172 GPa and is close to the experimentally available $T_c$ value of 25 K at 161 GPa. As pressure increases, the *d*-bands near the Fermi surface is found to shift downward to points where valence electrons in Ca transfer from *s*-bands to *d*-bands. Such transfer of charges at ~172 GPa may be thought to be completed, where the calculated $T_c$ (24.7 K) is found to be maximum.

**Acknowledgments**

The authors acknowledge the help received from Rajshahi University and Bangladesh University Grants Commission.